%% file: paper.tex
\documentclass[sigconf, screen, nonacm]{acmart}

\AtBeginDocument{%
  }

\setcopyright{none} 

\usepackage{amsmath,amsfonts}
\usepackage{algorithmic}
\usepackage{algorithm}
\usepackage{array}
\usepackage{graphicx}
\usepackage[caption=false,font=footnotesize,labelfont=sf,textfont=sf]{subfig}
\usepackage{textcomp}
\usepackage{stfloats}
\usepackage{url}
\usepackage{verbatim}
\usepackage{dsfont}

\usepackage{draftwatermark}

\SetWatermarkText{PRE-PRINT}

\SetWatermarkColor[gray]{0.8}

\SetWatermarkAngle{45}

\SetWatermarkScale{0.8}

\begin{document}
\title{SEADA: An efficient methodology for optimizing mixed-precision DNNs on multi-precision spatial architectures
}

\author{Leandro Fiorin}
\affiliation{%
  \institution{Indipendent Researcher}
  \city{Groningen}
  \country{The Netherlands}
}
\authornote{This work was performed while at Politecnico di Milano.}
\email{leandro.fiorin@gmail.com}
\authornote{Corresponding author}

\author{Marco Ronzani}
\affiliation{%
  \institution{Dipartimento di Elettronica e Informazione, Politecnico di Milano.}
  \city{Milano}
  \country{Italy}
}
\email{marco.ronzani@polimi.it}

\author{Cristina Silvano}
\affiliation{%
  \institution{Dipartimento di Elettronica e Informazione, Politecnico di Milano.}
  \city{Milano}
  \country{Italy}
}
\email{cristina.silvano@polimi.it}

\begin{abstract}
Mixed-precision computation has been introduced in deep neural networks (DNNs) as an effective approach to reduce latency, energy consumption, and memory footprint. 
However, efficiently mapping mixed-precision networks onto multi-precision spatial architectures poses several challenges. 
These include determining the appropriate precision for each layer, balancing layer-wise accuracy sensitivity to quantization against architectural heterogeneity and system-level constraints, and accurately estimating the system-level cost of heterogeneous precision assignments.

This work presents SEADA, an efficient methodology designed to address these challenges. 
SEADA comprises:
(i)  a configurable system-level analytical cost model of a multi-precision spatial accelerator architecture;
(ii) a fast mapping tool that identifies near-optimal mappings of DNN workloads onto the target integer accelerator;
(iii) analytical models for floating-point layers to estimate the overall benefits of mixed-precision execution;
and (iv) a per-layer precision selection methodology based on bit-level entropy, enabling efficient assignment across multiple numerical precisions.

SEADA's efficiency provides designers with a robust framework for the design-space exploration of multi-precision architectures\footnote{In this work, mixed-precision refers to DNN models with layer-wise precision variations, while multi-precision denotes hardware support for multiple precisions within the accelerator.}.
\end{abstract}

\maketitle

\input{sections/introduction}
\input{sections/background}
\input{sections/seada}
\input{sections/experiments}
\input{sections/dse}
\input{sections/related}
\input{sections/conclusions}
\bibliographystyle{ACM-Reference-Format}
\bibliography{biblio}
\end{document}

%% file: sections/introduction.tex
\section{Introduction}
\label{sec:introduction}
The demanding energy, performance, and memory requirements of DNN workloads have driven the development of highly optimized hardware accelerators and memory subsystems.
DNN training is commonly performed on GPUs with floating-point (FP) arithmetic, while inference is frequently mapped to energy-efficient hardware accelerators that employ spatial architectures and limited-precision quantized operations to minimize computation costs\cite{Silvano-2025}.
The observation that different layers of a DNN do not require the same numerical precision to accurately represent information has motivated the investigation of mixed-precision systems, in which each layer can potentially operate at a different precision \cite{Yao-2021, Bablani-2024}.
Efficiently mapping a mixed-precision network onto a multi-precision spatial architecture introduces several challenges.
In particular, determining the appropriate precision for each DNN layer is a non-trivial optimization problem, as it must balance layer-wise accuracy sensitivity to quantization with architectural heterogeneity and system-level constraints, while minimizing the overall accuracy degradation. 
Additional challenges include coordinating heterogeneous compute units, managing precision-specific data paths and memory accesses, and mitigating 
overheads arising from precision conversions and workload imbalance \cite{Gholami-2021}.

As illustrated in Figure~\ref{fig:expl}, a conventional state-of-the-art workflow for addressing this optimization problem requires three primary inputs: the neural network characteristics, the target training dataset, and a performance model of the spatial architecture~\cite{Fiorin-2025, Klhufek-2024}. 
Based on these parameters, the workflow selects an optimal candidate distribution for layer-wise numerical precisions.
The resulting quantized network is then mapped onto the accelerator model and evaluated with respect to the chosen optimization objective (e.g., number of bit multiply–accumulate operations (bMACs), energy consumption, or execution latency).
In parallel, the quantized network is trained for a fixed number of epochs to assess its inference accuracy. 
The measured accuracy, together with the corresponding cost metrics, is used to construct the Pareto frontier that characterizes the explored design space. 
If the termination criterion is not met, a new layer-wise precision configuration is selected and the process is iterated until convergence.

For a DNN with $L$ layers and a pool of $B$ candidate precisions per layer, the total number of possible layer-wise precision configurations is $B^L$.
Each of these configurations must be potentially trained to evaluate their accuracy and mapped onto the hardware accelerator to evaluate the selected optimization metric.
The mapping space itself further depends on the architecture’s parallelism and the dimensions of each layer. 
For example, for a ResNet-18\cite{He-2016}, the size of the combined design and mapping space is on the order of $\mathcal{O}(10^{33})$ \cite{Kao-2020}, making exhaustive exploration practically impossible.

Recent research into mixed-precision optimization has explored various methodologies, including integer linear programming and heuristic searches, to determine optimal layer-wise numerical precisions \cite{Yao-2021, Bablani-2024, Chen-2021, Dong-2020}.
However, these approaches often rely on high-level proxy metrics (like bMAC counts) that ignore critical architectural factors such as data movement costs, microarchitectural constraints, and mapping strategies.
Furthermore, existing analyses frequently overlook the performance and energy overhead of (de-)quantization when low-precision layers interface with full-precision components, and the actual benefits of mixed-precision execution are often overestimated \cite{Lin-2025}.

Complementary research optimizes DNN mapping and scheduling on spatial accelerators using brute-force, mixed-integer linear programming, or heuristics~\cite{Kwon-2019, Kao-2020, Symons-2021, Huang-2021, Jung-2023, Ronzani-2025.2, Parashar-2019, Klhufek-2024, Inci-2023}. 
However, these methods typically assume a fixed numerical precision and do not directly extend to multi-precision architectures, where precision heterogeneity fundamentally alters performance and data movement trade-offs.

Despite prior efforts to tackle these challenges in an integrated manner~\cite{Fiorin-2025, Klhufek-2024}, there is currently a lack of a unified and comprehensive framework to address the problem in its entirety.

This work presents SEADA, an efficient methodology that provides a comprehensive solution to the limitations of the current state-of-the-art.
Given a specific computational budget, SEADA rapidly estimates the optimal per-layer precision configuration and its associated accuracy degradation.
The computational cost is determined via analytical estimates, derived from the optimal mapping of mixed-precision layers onto a target multi-precision spatial accelerator.
With SEADA, we extend the state-of-the-art in several key directions:

\begin{itemize}

\item  We propose a template architecture for a multi-precision accelerator, enabling the analytical evaluation of execution trade-offs for mixed-precision kernels;

\item We develop analytical models for floating-point operations to estimate the overall benefits of mixed-precision execution.
We demonstrate that quantization overheads can be neutralized by fusing these operations directly into the coefficients of subsequent computational steps;

\item We develop a mapping tool capable of efficiently mapping mixed-precision kernels onto multi-precision spatial architectures, explicitly accounting for operands' precision in computation and data movement;

\item We extend an existing per-layer precision selection methodology based on bit-level entropy enabling fast precision assignment that requires only a single fine-tuning phase of the resulting mixed-precision network.

\end{itemize}

These contributions enable SEADA to efficiently address the target optimization problem, providing designers with a robust framework for the design-space exploration of multi-precision architectures.
To demonstrate its utility, we present a holistic exploration study that jointly optimizes the accelerator memory hierarchy, precision selection, and objective metrics.
A key insight from this study is that once the DNN accuracy degradation is characterized relative to the layer-wise precision distribution, a simple metric based on the number of bMACs of the DNN model is sufficient to estimate hardware-level accuracy.
Consequently, we introduce the Computational Bit Reduction (CBR) metric, a hardware-agnostic proxy for accuracy degradation. 
Since CBR depends solely on layer entropy, it requires only a single evaluation across all explored architectural configurations.

This paper is organized as follows.
Section~\ref{sec:background} provides relevant background about quantization and mixed-precision computations.
Section~\ref{sec:seada} discusses the several components of SEADA.
Section~\ref{sec:experiments} presents experimental evaluations of the methodology, while Section~\ref{sec:dse} discusses its use in a design-exploration flow.
Section~\ref{sec:related} discusses related work, while Section~\ref{sec:conclusions} concludes the paper with final remarks.

%% file: sections/background.tex
\section{Background}
\label{sec:background}

\subsection{Quantization methods and computations}
When designing quantized neural network systems, several orthogonal design choices must be considered, each influencing numerical precision, hardware efficiency, and model accuracy.

Quantization can be \textit{uniform} or \textit{non-uniform}, \textit{symmetric} or \textit{asymmetric}, and \textit{static} or \textit{dynamic}.
Uniform quantization applies evenly spaced quantization levels across the value range, 
while non-uniform schemes allocate levels unevenly to better match data distributions at the cost of increased complexity.
Symmetric quantization centers the quantization grid around zero with identical positive and negative ranges, while asymmetric quantization introduces a non-zero offset to better capture skewed values.
Static quantization determines scale and zero-point parameters ahead of inference (e.g., from calibration data), whereas dynamic quantization computes them on the fly.
Depending on accuracy and computation simplicity goals, different choices can be adopted.
In the case of low-precision quantization, an optimal choice appears the adoption of \textit{uniform symmetric static quantization} for weights and \textit{uniform asymmetric static quantization} for activations, where the quantization ranges are predetermined by analyzing the weights in one case, and by analyzing the range of activations for different batches, in the latter case\cite{Bablani-2024, Esser-2020}.

Quantization constrains DNN operands to a discrete set of representable values, defined by a mapping of real numbers to integers.
This process can be expressed as in Equation~\ref{equ:quantization}, where $\mathrm{Q}(\cdot)$ denotes the quantization operator, $r$ is a real-valued weight or activation, $S_{r}$ is a real-valued scaling factor, and $z_{r}$ is the zero point selected to ensure that the real value 0 is mapped exactly to an integer, equal to zero in case of symmetric quantization.
The operator $\mathrm{Int}(\cdot)$ converts a floating-point value to an integer via a chosen rounding rule (e.g., round-to-nearest or truncation).

{\small
\begin{equation}
Q(r) = r_{q} = \mathrm{Int}\left(\frac{r}{S_{r}}\right) - z_{r}
\label{equ:quantization}
\end{equation}
}
De-quantization $\mathrm{DQ}(\cdot)$, reconstructs an approximate full-precision representation of the quantized value and can be expressed as:
{\small
\begin{equation}
r \approx DQ(r_{q}) = S_{r} \mathrm{FP}(r_{q} + z_{r})
\label{equ:de-quantization}
\end{equation}
}
where the operator $\mathrm{FP}(\cdot)$ converts an integer value into floating-point representation.
For the sake of clarity, the zero-point $z_r$ is neglected in the subsequent formulations.
Nevertheless, it is fully accounted for in the actual implementation of our analytical models.

The choice of the scaling parameter $S_{r}$ is commonly determined by analyzing the dynamic range of the tensor to be quantized, either by using the maximum absolute value in symmetric quantization or by computing the minimum and maximum values in asymmetric quantization.
In post-training quantization (PTQ) approaches, these statistics are typically collected from a representative calibration dataset, whereas in quantization-aware training (QAT) they may be gathered or refined during training\cite{Zhao-2023, Zhang-2025, Gholami-2021}.
These range estimates define the interval that is mapped onto the available discrete quantization levels.
For very low numerical precision, such methods often become suboptimal because they fail to adapt to the network's optimization dynamics and loss landscape.
In this regime, learned-quantizer methods, such as Learned Step Size Quantization (LSQ)\cite{Esser-2020}, provide a more effective quantization-aware training alternative.
By treating the scale parameter $S_{r}$ as a trainable variable for a specific weight group (typically per-layer or per-channel), these methods jointly optimize the quantizer alongside the network weights, 
allowing the model to adapt precisely to quantization sensitivity, significantly improving accuracy under aggressive low-bitwidth constraints.
Building on these advantages, our work adopts the LSQ approach, specifically utilizing \textit{channel-wise learned scale parameters for weights} and \textit{layer-wise parameters for activations}.
This hybrid configuration balances the granular precision needed to capture weight distribution variance across channels with the computational efficiency of a single scale factor for activation maps.

Quantized operands are mostly used in DNNs to reduce computation costs of kernels implementing vector-matrix or matrix-matrix multiplication.
Considering a layer with hidden activations $h$ and a weight tensor $W$, their multiplication (convolution) can be expressed in terms of their respective quantized values as $h_{q}$ and $W_{q}$: 
{\small
\begin{equation}
a = S_{W} S_{h} FP(W_{q} * h_{q})
\label{equ:a1}
\end{equation}
}
where $S_{h}$ and $S_{W}$ are the real-valued quantization scales of $h$ and $W$, respectively.
The term $W_{q} * h_{q}$ represents the matrix multiplication (or convolution) carried out entirely in low-precision integer arithmetic with accumulation performed in INT32.

This intermediate result $a$ can be subsequently re-quantized before being forwarded to the next layer:
{\small
\begin{equation}
a_{q} = \mathrm{Int}\left(\frac{a}{S_{a}}\right) = \mathrm{Int}\left(\frac{S_{W} S_{h}}{S_{a}}FP(W_{q} * h_{q})\right)
\label{equ:out_quant}
\end{equation}
}
where $S_{a}$ is the precomputed scaling factor for the output activations.

\subsection{Layer precision selection}
The basic idea behind the use of mixed-precision quantization is based on the intuition that different layers in a network have different sensitivity to quantization and hence some layers are better candidates for more aggressive quantization than others\cite{Bablani-2024, Yao-2021, Chen-2021}.
Under the assumption that layer-wise accuracy estimates can be combined linearly to determine overall network accuracy\cite{Bablani-2024}, the problem of selecting the optimal precision for each layer can be formulated either as maximizing the total network accuracy or minimizing the accuracy perturbation due to the use of lower precision layers, while constraining computational costs (e.g., latency, number of bMACs, energy).
The problem is usually solved using Integer Linear Programming (ILP), or by formulating the optimization as a Multiple-Choice Knapsack Problem\cite{Bablani-2024, Yao-2021, Chen-2021}.

Our methodology leverages the Entropy Approximation Guided Layer selection (EAGL) strategy proposed in \cite{Bablani-2024}.
Given a set of candidate precisions $B = \{b_{0}, b_{1}\}$, EAGL formulates the optimization as a 0-1 Knapsack Problem.
The objective is to maximize the cumulative accuracy gain $\sum_{l=1}^{L}G_lP_l$ subject to a total computational budget $T$:
$$\text{maximize} \sum_{l=1}^{L}G_lP_l \quad \text{subject to} \sum_{l=1}^{L}C_l \leq T$$
In this formulation, $G_l$ represents the estimated accuracy gain from maintaining layer $l$ at the higher precision $b_{1}$, while $P_l$ is a binary indicator variable ($P_l=1$ for high precision $b_{1}$, and $0$ otherwise). 
$C_l$ denotes the computational cost of layer $l$, and $T$ is the global budget constraint expressed as an integer.
To quantify $G_l$ without exhaustive search, EAGL uses empirical layer entropy as a proxy for quantization sensitivity. 
For a model pre-quantized at $b_0$, EAGL maps weights to $2^b$ discrete bins to evaluate the necessity of a higher precision, $b_1$. 
High entropy, resulting from weights distributed evenly across bins, indicates a benefit from increased precision. 
Conversely, low entropy signifies weight concentration, suggesting the layer's transformation can be represented more compactly at a lower bit-width.

Given the vector $w$ of $n$ trained parameters of layer $l$ quantized at precision $b$, the empirical entropy for each layer is computed as $H(\hat{p}_l^b) = - \sum_{c}\hat{p}_l^b(c) log (\hat{p}_l^b(c))$, where the empirical probability distribution for each discrete value $|c| \leq 2^b$ is computed as $\hat{p}_l^b(c) = \frac{1}{n} \sum \mathds{1}_c(w_i)$, with $\mathds{1}_c(w_i)$ equal to $1$ if $w_i = c$, and $0$ otherwise.

Unlike other approaches, EAGL requires only a single pre-trained model quantized at precision $b$ to estimate parameter distributions.
While this may occasionally yield suboptimal precision assignments, it is orders of magnitude faster than alternative methods \cite{Bablani-2024}.
This efficiency makes EAGL an ideal candidate for integration into comprehensive estimation methodologies.

\subsection{Floating-point layers in DNN models}

\begin{figure}[t!]
 \centering
 	\subfloat[\label{fig:bert_cm}]{
	   \includegraphics[width=0.95\linewidth]{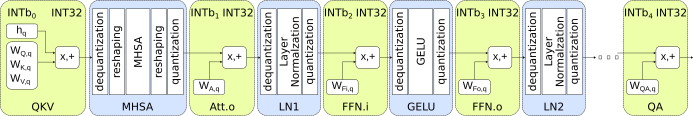}
    }
    \\  
    \subfloat[\label{fig:resnet50_cm}]{
	   \includegraphics[width=0.95\linewidth]{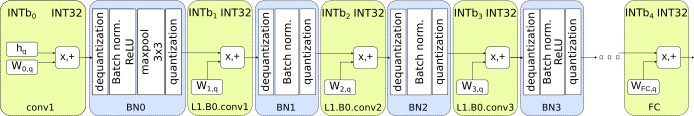}
    }
   \caption{DNN computational models: (a) BERT-base for question answering and (b) ResNet-50. Green boxes highlight computations executed using low-precision integer operands, while blue boxes denote computations performed in full floating-point precision.}
  \label{fig:dnn_cm}
  \vspace{-0.5 cm}
\end{figure}

Figure~\ref{fig:dnn_cm} outlines the comprehensive computational pipelines of the two DNN models selected in this work as use-cases: a BERT-base language model fine tuned for question-answering \cite{Devlin-2018}, and a ResNet-50 CNN \cite{He-2016}. 
For clarity, the figure focuses on the sequential processing stages and omits residual (skip) connections.

The BERT-base architecture consists of twelve transformer blocks.
Each block sequentially computes the Query, Key, and Value matrices (QKV), followed by Multi-Head Self-Attention (MHSA) and an attention output projection (Att.o).
These operations are supported by Layer Normalization (LayerNorm - LN1, LN2), intermediate and output Feed-Forward Networks (FFN.i, FFN.o), and Gaussian Error Linear Unit (GELU) activation functions.
The pipeline concludes with a specialized layer for the question-answering (QA) task.

The ResNet-50 model features a backbone of interleaved convolutional and Batch Normalization (BatchNorm - BN) layers, culminating in a final Fully Connected (FC) layer for classification.

In Figure~\ref{fig:dnn_cm}, green boxes denote low-precision integer operations offloaded to accelerators, while blue boxes represent full-precision floating-point steps.
This \textit{simulated quantization} approach \cite{Kim-2021} stores parameters as integers but performs only a subset of arithmetic in the integer domain.
Consequently, data must be de-quantized for floating-point operations and re-quantized for subsequent low-precision stages.

Prior research has explored uniform integer execution for BERT \cite{Kim-2021} and fully integer mixed-precision for CNNs \cite{Yao-2021}. 
However, the ability of these techniques to maintain accuracy under very low-precision constraints remains unproven. 
Consequently, this work focuses on simulated quantization, leaving fully integer mixed-precision models for future investigation.

%% file: sections/seada.tex
\section{SEADA}
\label{sec:seada}

\begin{figure}[!t]
    \centering
    \subfloat[\label{fig:expl}]{%
        \includegraphics[width=0.3\textwidth]{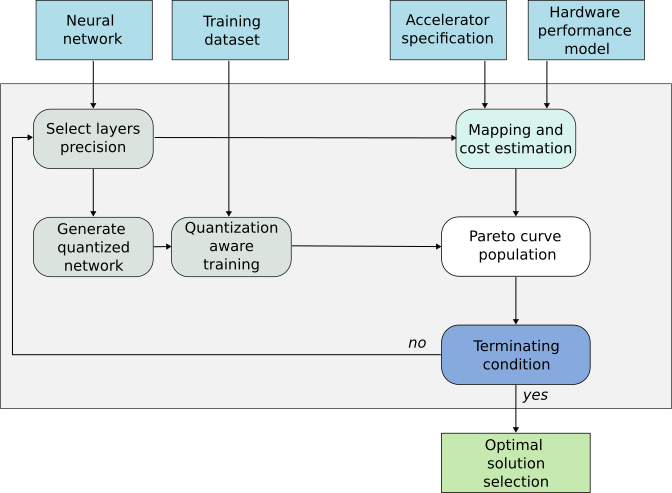}
    }
    \hfill
    \subfloat[\label{fig:seada}]{%
        \includegraphics[width=0.3\textwidth]{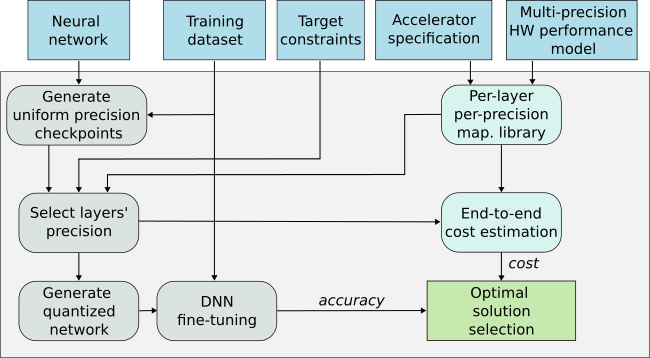}
    }
    \caption{(a) Overview of a state-of-the-art  framework for the joint optimization of hardware performance and inference accuracy in DNN spatial accelerators; (b) Overview of the main components of the proposed SEADA methodology.}
    \label{fig:tools}
    \vspace{-0.5 cm}
\end{figure}

This section describes SEADA, a methodology for efficiently exploring the design space of mixed-precision neural networks deployed on multi-precision accelerators. 
SEADA integrates several key components: (i) a configurable system-level analytical cost model of a multi-precision spatial accelerator architecture; (ii) a fast mapping tool that identifies near-optimal mappings of DNN workloads onto the target integer accelerator; (iii) analytical models for floating-point layers to estimate the overall benefits of mixed-precision execution; and (iv)a per-layer precision selection methodology based on bit-level entropy, enabling efficient assignment across multiple numerical precisions.

Figure~\ref{fig:seada} illustrates the interaction among the different components of the proposed methodology.
An EAGL-based approach is employed to determine the per-layer precision configuration of the quantized network. 
The precision selection process takes as inputs: (i) a library of cost estimates for each network layer and precision, generated by the mapping tool; (ii) the allowed cost budget; and (iii) entropy measures computed from checkpoints obtained by pre-training the network at uniform precisions.

Once the per-layer precision configuration is determined, the resulting mixed-precision network is fine tuned to evaluate the corresponding accuracy, while the comprehensive computational cost is estimated using the analytical model.
This approach yields an optimal or near-optimal solution without requiring iterative exploration over multiple per-layer precision configurations, in contrast to the iterative workflow illustrated in Figure~\ref{fig:expl}.

We detail in the following subsections the main SEADA components.

\input{sections/accelerator}

\subsection{Mapping tool for convolution and GEMM}
\label{ssec:cost-model}

We leverage the \textit{QuickFlow} mapping tool~\cite{Ronzani-2025.2} to evaluate both performance and cost of mixed-precision convolutional and general matrix–matrix multiplication (GEMM) kernels.

QuickFlow optimizes DNN mappings (convolutions and GEMM) onto hierarchical spatial accelerators using an iterative local search.
By evaluating candidates through an analytical, activity-based cost model, validated against Timeloop~\cite{Ronzani-2025.1, Parashar-2019} and integrated with Accelergy~\cite{Wu-2019}, it navigates tilings and dataflows to minimize latency and energy. 
It represents architectures as hierarchical trees, from leaf-level arithmetic units to DRAM, allowing for automated topology inference and optimal resource scheduling.
QuickFlow's mapping exploration typically assumes a uniform precision for all computations and data transfers within a single run. 
While the tool can optimize individual layers at different precisions, it treats each as a distinct architectural baseline. 
Consequently, it cannot accurately model hardware flexibility or the specific performance characteristics of multi-precision processing elements operating within a unified system.

To address these limitations, we extended the baseline QuickFlow implementation by explicitly modeling the $\sum_{i} A_i \cdot B_i + C$ operation executed at reduced precision as a cluster of $P$ FMA units, with final accumulation performed at INT32 precision.
Spatial mapping constraints are imposed on this cluster, such that each low-precision functional unit is restricted from accumulating partial sums corresponding to different output feature-map locations or different output channels.
Supporting multi-precision execution further required extending the modeling of precision-dependent adder reduction trees and partial-result accumulation, as well as accurately capturing the packing and unpacking of input and output data as they are transferred across the different levels of the memory hierarchy.

\subsection{Floating-point layer models}
\label{ssec:fp-model}
To holistically evaluate mixed-precision execution, we developed activity-based analytical models for the floating-point stages (blue boxes, Figure~\ref{fig:dnn_cm}).
These models estimate component-level activity and scale it by the cost per operation, following established methodologies~\cite{Ronzani-2025.2, Parashar-2019, Wu-2019}.

For BERT-base (MHSA, GELU, LayerNorm) and ResNet-50 (BatchNorm), floating-point stages often utilize Special Function Units (SFUs) for complex operations like exponentiation, division, and trigonometric functions.
While precise cost estimation requires specific hardware and kernel-level details, we approximate these costs by decomposing the kernels into equivalent sequences of simpler floating-point operations, following state-of-the-art algorithms and approximations commonly used in optimized DNN libraries~\cite{Wang-2025, Andri-2025}.

In our evaluation, we analytically demonstrate that the overhead associated with implementing the quantization and de-quantization of accelerated layers can be minimized by fusing these operations into the coefficients of subsequent floating-point computation steps.
An additional, albeit simple, yet important observation is that the computational cost of these fused kernels marginally depend on the numerical precision used in the preceding and subsequent layer. 
Consequently, the accelerated layers can be optimized independently of these operations.

\subsubsection{Multi-Head Self-Attention}

The Multi-Head Self-Attention (MHSA) layers in BERT (Figure~\ref{fig:bert_cm}) capture contextual dependencies between all tokens in a sequence. 
By operating multiple heads in parallel, the model explores diverse representational subspaces to identify various relationship types. 
The resulting aggregated outputs provide the context-aware representations essential for complex language understanding.
By fusing the de-quantization and quantization steps directly into the MHSA computation, the resulting operation can be expressed as:
{\small
\begin{equation}
\begin{split}  
& a_{q}^{MHSA} = Q(SM(\frac{Q_{m} * K_{m}^{T}}{\sqrt(d_{m})}) * V_{m}) \\
& = \mathrm{Int}\left(SM(\frac{S_{Q} S_{K}}{\sqrt(d_{m})}FP(Q_{q}) * FP(K^T_{q})) * \frac{S_{V}}{S_{a}} FP(V_{q})\right) \\
&= \mathrm{Int}\left(SM(\omega_{0} FP(Q_{q}) * FP(K^T_{q})) * \omega_{1}FP(V_{q})\right)
\end{split}
\label{equ:dequan-softmax-quan}
\end{equation}
}

where $SM(\cdot)$ denotes the \textit{softmax} function, $Q_{m}$, $K_{m}$, and $V_{m}$ denote the per-head partitions of the Query, Key, and Value matrices, respectively, and $d_{m}$ is the corresponding scaling factor.
The parameters $\omega_{0}$ and $\omega_{1}$ are precomputed constants, while all remaining terms follow the notation introduced previously.

The computational efficiency of MHSA is primarily limited by the memory bandwidth required for large intermediate matrices.
FlashAttention-2~\cite{Dao-2023} mitigates this by using tiled access to $Q, K,$ and $V$ matrices and fusing attention operations into a single kernel.
By implementing an online softmax, it avoids materializing intermediate scores, reducing memory traffic and improving memory complexity from quadratic to linear relative to sequence length.

We adapted the FlashAttention-2 algorithm to the specific constraints of our target architecture and parallelized the computation across the DNN sequence length, assigning each processing element to execute the fused attention kernel for a single input token.
This approach enables an analytical model that precisely captures memory access patterns across the hierarchy, accounting for data movement between local registers, shared memory, and global storage.

To evaluate the cost of the fused kernel, expressed in our case as $softmax(\omega_{0} FP(Q_{q}) * FP(K^T_{q})) * \omega_{1} FP(V_{q})$, we derive the number and types of floating-point operations required to implement the online softmax computation described in~\cite{Dao-2023}. 
Specifically, our analysis accounts for the polynomial approximation of the exponential function proposed in~\cite{Belano-2025}, as well as three iterations of the Newton–Raphson method to compute the reciprocal of the softmax normalization term using only multiplications and additions~\cite{Press-2007}.

\subsubsection{GELU}
The Gaussian Error Linear Unit (GELU) layer in the BERT-base model serves as a smooth, non-monotonic activation function.
It introduces non-linearity by scaling the input based on its cumulative distribution function.
The GELU approximation employed in Hugging Face’s BERT implementation is expressed as $GELU(x) = k_{0} x (1+tanh(k_{1}(x+k_{2}x^3))$, with $k_{0} = 0.5$, $k_{1} = \sqrt(2/\pi)$, and $k_{2} = 0.044715$ \cite{Wolf-2020}.
By fusing de-quantization, GELU, and quantization into a single expression, the resulting operation can be written as:

{\small
\begin{equation}
\begin{split}  
& a_{q}^{GELU} = Q(GELU(DQ(h_{q})) \\
& = \mathrm{Int}\left(k_{0}^{'}FP(h_{q})(1+tanh (k_{1}^{'}(FP(h_{q}) + k_{2}^{'}FP(h_{q})^3)\right)
\end{split}
\label{equ:dequan-GELU-quan}
\end{equation}
}

where $k_{0}^{'} = k_{0} S_{h}/S_{a}$, $k_{1}^{'} = k_{1}S_{h}$, and $k_{2}^{'} = k_{2}S_{h}^{2}$.

Since $\tanh(z)$ can be expressed as $\tanh(z) = 1 - 2/(\exp(2z) + 1)$, we model the analytical cost of the GELU implementation by leveraging the same polynomial approximation of the exponential function adopted for the softmax computation~\cite{Belano-2025}, together with three iterations of the Newton–Raphson method to compute the reciprocal of the denominator\cite{Press-2007}.

The memory access patterns associated to the execution of the GELU function are straightforward to derive, as it is applied point-wise on each activation value, while the parameters $k_0'$, $k_1'$, and $k_2'$ are precomputed constants shared across all activations.

\subsubsection{Layer Normalization}
Layer Normalization (LayerNorm) stabilizes hidden state dynamics by normalizing activations across the feature dimension for each sequence element. 
By computing the mean $\mu$ and variance $\sigma^2$ independently for each token’s hidden tensor across dimension $H$, it ensures consistent normalization regardless of batch size or sequence length. This facilitates smoother gradient flow and significantly accelerates pre-training convergence.
The LayerNorm is typically formulated as:

{\small
\begin{equation}
y = \frac{h - \mu}{\sqrt{\sigma^2 + \epsilon}} \odot \gamma + \beta
\label{equ:layernorm}
\end{equation}
}
where $\gamma$ and $\beta$ are vectors of dimension $H$ of learnable parameters that are shared with by each token of the sequence length and batch.

The computation of LayerNorm is typically carried out in two stages.
First, Welford’s online algorithm\cite{Welford-1962} is used to compute the running mean and sum of squared deviations in a single pass over the input data. 
Second, the remaining normalization, scaling, and shifting operations are fused into a single computation.
When de-quantization, LayerNorm, and quantization are fused, the resulting expressions for the mean and variance can be written as:

{\small
\begin{equation}
\mu \approx \frac{1}{H} \sum_{i=1}^{H} S_{h}FP(h_{q}) = S_{h} \mu^{'} 
\label{equ:mu_prime}
\end{equation}
}
{\small
\begin{equation}
\sigma^2 \approx \frac{1}{H} \sum_{i=1}^{H} (S_{h} FP(h_{q}) - S_{h}\mu^{'})^2 = S_{h}^2 \sigma^{'2}
\label{equ:sigma_prime}
\end{equation}
}

The overall fused operation can be described by the formula:
{\small
\begin{equation}
\begin{split}
a_{q}^{LN} & = \mathrm{Int}\left(\frac{S_{h} FP(h_{q}) - S_{h}\mu^{'}}{\sqrt{S_{h}^{2}\sigma^{'2} + \epsilon}} \odot \frac{\gamma}{S_{a}} + \frac{\beta}{S_{a}}\right) \\
& = \mathrm{Int}\left(\frac{FP(h_{q}) - \mu^{'}}{\sqrt{\sigma^{'2} + \epsilon^{'}}} \odot \gamma^{'} + \beta^{'}\right)
\end{split}
\label{equ:layernorm_dequ}
\end{equation}
}
where $\gamma^{'} = \gamma/S_{a}$, $\beta^{'} = \beta/S_{a}$, and $\epsilon^{'}$ can be chosen small enough to satisfy the floating-point precision requirements to ensure numerical stability.

To model the cost of the LayerNorm, we consider the same two computing steps used in optimized libraries and parallelization along the sequence length's tokens.
The normalization requires computing the reciprocal of the modified standard deviation, which we approximate using three iterations of the Newton–Raphson method \cite{Press-2007}.

Since the second stage is performed point-wise on the activation data and shares the precomputed scaling and bias terms $\gamma'$ and $\beta'$, the corresponding memory access patterns are straightforward to derive.

\subsubsection{Batch Normalization}
Batch Normalization (BatchNorm) stabilizes and accelerates CNN's training by normalizing layer inputs to mitigate Internal Covariate Shift, preventing saturation and allowing for higher learning rates\cite{He-2016}.
Taking Figure~\ref{fig:resnet50_cm} as reference, the de-quantization, batch normalization, and quantization steps of a quantized hidden tensor $h_{q}$ can be fused into a modified BatchNorm layer, as described by Equation~\ref{equ:dequan-bn-quan}:

{\small
\begin{equation}
\begin{split}
a_{q}^{BN} & = Q(BN(DQ(h_{q})))
= \mathrm{Int}\left(\frac{BN(S_{h}FP(h_{q}))}{S_{a}}\right) \\
& = \mathrm{Int}\left(\gamma(\frac{S_{h} FP(h_{q}) - \mu}{S_{a}\sqrt{\sigma^{2} + \epsilon}}) + \frac{\beta}{S_{a}}\right) = \mathrm{Int}\left(\alpha_0FP(h_{q}) + \alpha_1\right)
\end{split}
\label{equ:dequan-bn-quan}
\end{equation}
}
%
where $BN(\cdot)$ implements BatchNorm.
The parameters $\mu$ and $\sigma^{2}$ represent the mean and variance computed during training along the batch and the spatial dimensions, while $\beta$ and $\gamma$ are the trainable affine parameters of the BatchNorm layer.
By abstracting away data type conversions, the operation expressed in Equation~\ref{equ:dequan-bn-quan} can be modeled as a sequence of point-wise FMA operations between the activation values and the $\alpha_{0}$ and $\alpha_{1}$ coefficients, which are constant and precomputed for all elements within the same batch.
This computation is readily parallelizable across the floating-point processing elements.
However, it exhibits low arithmetic intensity, since activation values are consumed only once with no reuse.

\subsection{Layer-wise precision selection}
SEADA aims to find an optimal precision distribution that maximizes model accuracy within specific computational, memory, or energy constraints. 
In mixed-precision settings, the combinatorial expansion of the search space often necessitates meta-heuristics like Genetic Algorithms or Reinforcement Learning~\cite{Fiorin-2025, Klhufek-2024}.
However, these methods are frequently computationally intractable for large-scale architectures, as the cumulative overhead of evaluating the accuracy of each candidate, even with shortened fine-tuning, is prohibitively expensive.

To navigate the design space efficiently, we implement an exploration strategy based on the EAGL approach~\cite{Bablani-2024}. 
We quantify computational overhead using the analytical cost model from Section~\ref{ssec:cost-model}, focusing exclusively on quantized layers since floating-point costs remain mostly constant across all configurations.
To maximize efficiency, we pre-compute metrics for all possible layer-wise precisions into a library of optimal mappings.
This allows SEADA to rapidly estimate total costs via look-up table aggregation, decoupling the search process from model execution.

We extended the original EAGL methodology to be able to select from more than two precisions by using a step approach.
Taking as example the case of three precisions $b_0$, $b_1$, and $b_2$, with $b_0 > b_1 > b_2$, we evaluate the layer entropy for a network trained with $b_0$ and use the obtained values to selected the layers to which to assign the precision $b_1$ until the computational budget is equal to the one obtained using $b_1$ for all layers.
For lower computational budget, we use instead as reference entropy the one computed for a network trained at precision $b_1$.

We also evaluated an alternative approach by modeling the precision assignment problem as a Multi-Choice Knapsack Problem, jointly considering the layer entropies for precisions $b_0$ and $b_1$. 
However, this method yielded inferior results, particularly under less restrictive computational budgets (results omitted for brevity).

Given the precomputed metrics, a bucket of precisions to be chosen from, and the computed layer entropy, SEADA automates the bit-width allocation by treating each layer's precision as a selectable item with an associated cost (the precomputed metric) and expected benefit (the entropy, as proxy for the accuracy). 
Unlike exhaustive search methods, this approach provides a deterministic and fast solution to the precision-assignment problem. 

To ensure consistency, activation tensors shared across multiple layers constrain those layers to a uniform precision. 
These layers are grouped into a single logical unit, where the total computational cost and accuracy gain are the aggregate sums of its constituent layers' metrics.

After finalizing the optimal configuration, we generate a mixed-precision PyTorch model by automatically modifying pre-trained architectures from repositories like Torchvision or Hugging Face.
This involves a layer-wise substitution where standard linear and convolutional layers are replaced with specialized modules utilizing Learned Step Size Quantization.

We estimate the final accuracy of the generated mixed-precision model by fine-tuning it using Knowledge Distillation~\cite{Hinton-2015} with the original full-precision model acting as a teacher network to guide the optimization of the quantized student model.
This process ensures high fidelity to the original performance, even under significant bit-depth constraints.

%% file: sections/accelerator.tex
\subsection{Multi-precision performance model}
\label{ssec:accelerator}
\begin{figure}[t]
 \centering
 \includegraphics[width=\columnwidth]{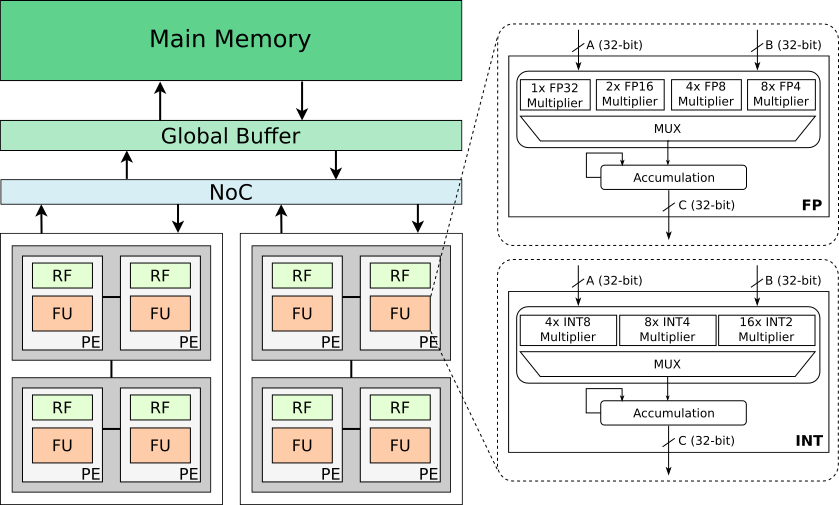}
 \caption{Target spatial accelerator architecture template.}
  \label{fig:architecture}
  \vspace{-0.5cm}
\end{figure}

To estimate the cost of implementing mixed-precision quantization, we introduced a system-level analytical model of a multi-precision spatial accelerator template designed for the execution of DNN workloads. 
The proposed model is grounded in existing accelerator architectures 
and abstracts key architectural features relevant to performance and energy analysis\cite{Chen-2016, Andersch-2022, Lee-2021, Kwon-2019}.

Figure~\ref{fig:architecture} illustrates the generic spatial accelerator template. 
The architecture comprises an array of processing elements (PEs), each equipped with local register files (RF) for storing input and output operands, a global buffer (GB), off-chip main memory, and an on-chip network (NoC) that interconnects all components.
The accelerator buffers are assumed to support double buffering to enable overlap of computation and data movement.
Additionally, PEs can be organized into multiple levels of physical clustering to expose different degrees of spatial parallelism.

Each PE features a configurable functional unit (FU) supporting mixed-precision computation, based on state-of-the-art designs~\cite{Lee-2021, Andersch-2022, Harris-2016}.
Following~\cite{Lee-2021}, the FU assumes decoupled pipelines for floating-point and fixed-point arithmetic, which can be independently optimized for energy efficiency.
While Figure~\ref{fig:architecture} shows both pipelines within each PE, our analytical model is flexible and easily adapted to alternative configurations, such as shared or centralized resources.

We assume a functional unit with 32-bit-wide input and output ports.
For the floating-point pipeline, we model support for FP32, FP16, FP8, and FP4 arithmetic\cite{Kalamkar-2019, Kuzmin-2022, Sami-2024, Xiao-2020}.
For the fixed-point pipeline, we incorporated models for computation in INT8, INT4, and INT2\cite{Lee-2021, Galal-2011}, with INT32 accumulation.

Both pipelines support fused multiply-add instructions (FMA).
In the case of the floating-point pipeline, when selecting 32 bit precision, the engine performs a single FMA.
When using a smaller quantization $Q$, the 32-bit lane is partitioned in $P = 32/Q$ operands, and the engine performs a $\sum_{P}^{} A_{i} \cdot B_{i} + C$ instruction.
Similarly, in the case of the fixed-point pipeline, the processing element can be programmed to perform any between four INT8 FMAs, eight INT4 FMAs, or sixteen INT2 FMAs in parallel, while accumulating computation results in the INT32 format.
We consider the multipliers of the floating and fixed point engines to be implemented with different numbers of pipeline stages~\cite{Lee-2021}.
As the targeted algorithms are significantly parallel, threads can be scheduled to minimize the influence of data dependency and achieve, ideally, a throughput of one instruction per cycle~\cite{Galal-2011}.
The template components are automatically instantiated depending on the precision requirements of the use cases targeted during the design-space exploration, together with the proper adder tree, accumulator data width and input and output operands packing.

As DNN workloads require more than just FMA operations, our floating-point pipeline models a broader instruction set, including basic arithmetic and data type conversions (omitted from Figure~\ref{fig:architecture} for simplicity)~\cite{Galal-2011, Wang-2025}.

%% file: sections/experiments.tex
\section{SEADA evaluation}
\label{sec:experiments}
This section evaluates the proposed SEADA methodology. 
As noted in Section~\ref{sec:related}, a direct comparison with state-of-the-art solutions is precluded by the absence of prior works that address the optimization problem holistically.
Furthermore, reproducing previous experiments is hindered by a lack of access to specific training hyperparameters and DNN model configurations used in those studies.

\subsection{Experimental setup}

\subsubsection{Reference spatial architecture}
In our experiments, we adopted the Eyeriss spatial  architecture~\cite{Chen-2016} as a reference platform, from which we derived system-level architectural characteristics, including the memory hierarchy, the number of PEs, and the basic dataflow.
Following~\cite{Lee-2021}, each PE is composed of both an integer pipeline and a floating-point pipeline. 
The multi-precision functional unit within the integer pipeline was modeled as described in Section~\ref{ssec:accelerator}.
We configured 168 PEs, each equipped with independent 1.66 KB register file sets, covering inputs, weights, and outputs, for both the integer and floating-point pipelines. The system also includes a 128 KB Global Buffer and LPDDR4 DRAM.
Latency and energy models for the integer pipeline, memory components, and interconnection network were generated using Accelergy~\cite{Wu-2019}, in a manner consistent with the methodologies adopted by QuickFlow and Timeloop~\cite{Parashar-2019}, and assuming a 32nm technology node and a clock rate of 1 GHz.
For our experiments, the floating-point pipeline was limited to FP32 operands, consistent with the accuracy evaluation setup.
Energy, performance, and area characteristics of the floating-point components were similarly derived using Accelergy.

For the low-precision kernels, we created the cost library by running the mapper with the objective of minimizing for each precision the layers' energy-delay product (EDP).
The execution of the floating-point kernels were parallelized over the available PEs, as described in Section~\ref{ssec:fp-model}.
In particular, as the efficiency of the Multi-Head Self-Attention strictly depends on the number of available PEs and on the size of the matrix tiles used in the computation of the FlashAttention-2 algorithm, we performed a exhaustive design-space exploration to find the configuration minimizing the EDP of the layer.

\subsubsection{DNN models pre-training}
We evaluated two representative benchmarks: a natural language question-answering task based on BERT-base~\cite{Devlin-2018}, and an image classification task based on ResNet-50~\cite{He-2016}.
The BERT-base model was obtained from the Hugging Face Transformers library~\cite{Wolf-2020} and fine tuned on the SQuAD 1.1 benchmark~\cite{Rajpurkar-2016}.
The ResNet-50 model was adapted from the PyTorch Model Zoo and trained on the ImageNet dataset~\cite{Deng-2009}. 

Both networks were first trained at full precision to serve as teacher models for the knowledge distillation procedure employed during the training of their low-precision counterparts. 
To compute the entropy metrics used for selecting the optimal precision distribution, we trained reference models in which all variable-precision layers were uniformly set to either 8 bits or 4 bits. 
Following common practice in quantized neural network design~\cite{Zhou-2016, Bablani-2024}, the first and last layers of each low-precision model were fixed at 8-bit precision.

We trained the BERT-base models for 60 epochs using a LAMB optimizer, an initial learning rate of $5\times10^{-5}$, a weight decay of $1\times10^{-2}$, and a batch size of 192.
The 8-bit model ResNet-50 was trained for 40 epochs using a SGD optimizer, a momentum of 0.9, a weight decay of $1\times10^{-4}$, an initial learning rate of $1\times10^{-3}$ with cosine learning-rate decay, and a batch size of 128.
For the 4-bit ResNet-50 model, we increased the initial learning rate to $1\times10^{-2}$, and used a weight decay of $2.5\times10^{-05}$.
All models were fine tuned using knowledge distillation, with a temperature of one and equal weighting between the standard task loss and the distillation loss.

\subsubsection{Mixed-precision models fine tuning}
The variable-precision layers of the mixed-precision models were allowed to operate at 8, 4, or 2 bits.
We initialized the mixed-precision models from the optimal 4-bit checkpoints and fine tuned them using the same training configuration adopted for the 4-bit models, for an additional maximum number of epochs equal to 60 for BERT, and to 40 for ResNet-50, and using three different seeds.
To provide a stable initialization for further training, the learned quantization scale factors were multiplied by four for layers whose precision was reduced from 4 bits to 2 bits, while they were left unchanged for layers operating at 8-bit or 4-bit precision.

For all experimental tasks, we evaluated the optimal layer-wise precision distributions by constraining the total cost metric to a fraction of the baseline configuration in which all quantized layers operate at 8-bit precision.
We conducted experiments using energy as the cost metric provided to the Knapsack-based optimization algorithm.
For BERT, we evaluated energy budgets ranging from 65\% down to 35\% of the 8-bit baseline in 5\% increments. 
For ResNet-50, we explored budgets from 75\% to 45\% of the baseline using 10\% steps.

\subsection{Experimental results}

\begin{figure*}[!t]
    \centering
    \subfloat[\label{fig:bert_accuracy}]{%
        \includegraphics[width=0.3\textwidth]{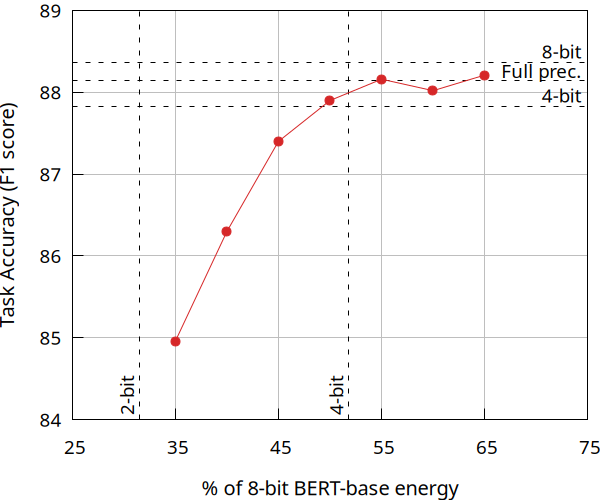}
    }
    \hfill
    \subfloat[\label{fig:bert_edp}]{%
        \includegraphics[width=0.3\textwidth]{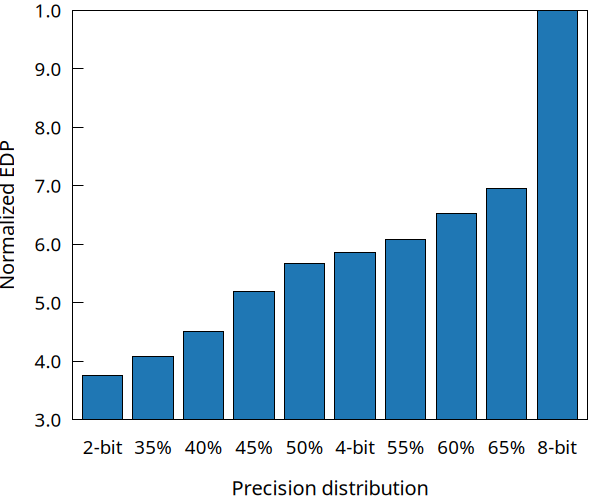}
    }
    \hfill
    \subfloat[\label{fig:bert_breakdown_energy_e2e}]{%
        \includegraphics[width=0.3\textwidth]{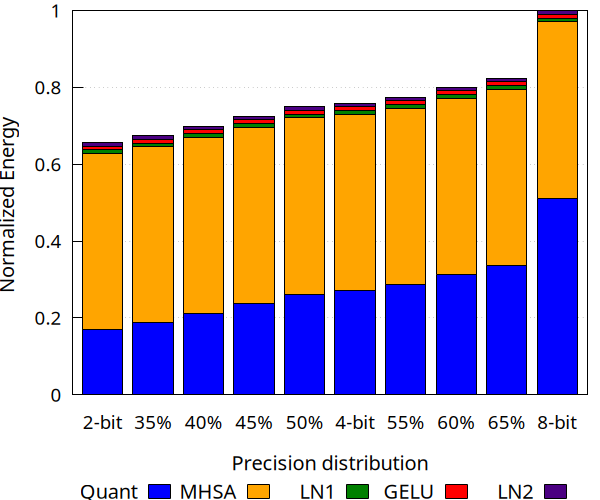}
    }
    \caption{BERT-base: (a) Achieved fined-tuned accuracy for different energy budgets with respect to an uniform 8-bit implementation of the accelerated kernels; (b) Normalized EDP of comprehensive model for different distributions of layers' precision; (c) Normalized comprehensive energy broken down in terms of the layer type.}
    \label{fig:bert}
\end{figure*}

\begin{figure}[t]
 \centering
 \includegraphics[width=0.7\columnwidth]{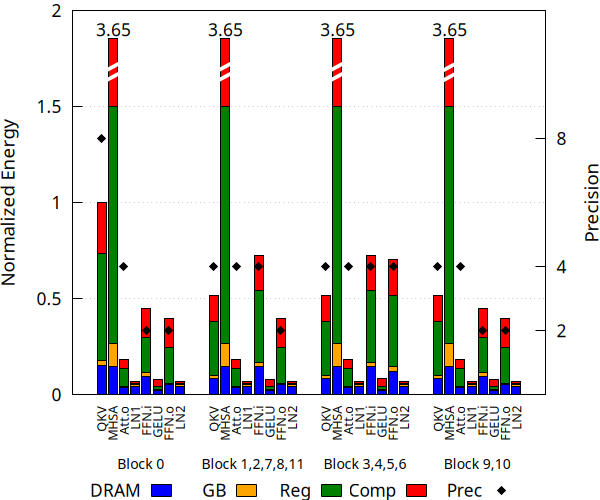}
 \caption{Normalized comprehensive energy broken down for architectural components, for BERT-base blocks and layers. 
 Results are normalized to the energy consumption of the initial QKV layers. Precision values are reported exclusively for the accelerated layers of the model.}
  \label{fig:bert_breakdown_energy_layer_grouped}
  \vspace{-0.5cm}
\end{figure}

\subsubsection{BERT-base}
\label{sec:res-bert}
Figure~\ref{fig:bert_accuracy} illustrates the accuracy versus energy-saving trade-offs achieved by applying our methodology to BERT-base, excluding the energy consumption of floating-point layers.
For energy budgets above 45\%, the accuracy loss remains within 1\% of the full-precision baseline; however, accuracy degrades significantly as the energy budget is further constrained.

Figure~\ref{fig:bert_edp} compares the normalized EDP across the comprehensive network model, including both accelerated and floating-point layers. 
The comparison evaluates layer acceleration at 2, 4, and 8 bits alongside the mixed-precision networks selected for the seven energy budgets.
The EDP reduction observed at lower precisions stems from decreased energy consumption, minimized data movement within accelerated kernels, and reduced latency during compute-intensive operations.
The EDP reduction with respect to the 8-bit configuration spans from 30\% for the 65\% mixed-precision solution to 59\% for the 35\% configuration.
Specifically, the 45\% mixed-precision solution achieves a 48\% EDP reduction relative to the 8-bit baseline and a 97\% reduction compared to the full-precision implementation.

Figure~\ref{fig:bert_breakdown_energy_e2e} illustrates the normalized energy consumption for various precision distributions, categorized by BERT layer types: accelerated low-precision matrix multiplications (Quant), MHSA, LayerNorm (LN), and GELU.
Once the compute kernels are quantized, the MHSA computation dominates the total energy profile, aligning with the FlashAttention-2 acceleration results reported in \cite{Dao-2023, Wang-2025}.
Overall energy savings with respect to the 8-bit configuration range from 17.7\% for the 65\% mixed-precision solution to 32.7\% for the 35\% configuration. 
Notably, the 45\% mixed-precision network achieves a 27.6\% energy reduction compared to the 8-bit baseline and an 88.4\% reduction compared to the full-precision implementation.
The reduced energy saving with respect to the case in which only the accelerated layers are considered underscores the necessity of a holistic approach to system-level optimization.

Figure~\ref{fig:bert_breakdown_energy_layer_grouped} illustrates the per-layer normalized energy breakdown for the 45\% mixed-precision network, categorized by the precision configurations selected for each of the twelve BERT blocks. 
The breakdown partitions energy consumption into main memory (DRAM), global buffer (GB), register files (Reg), and computational units (Comp).
The results highlight that the energy budget is primarily consumed by the computational units and register files of compute-bound layers, such as QKV, MHSA, Att.o, FFN.i, and FFN.o. 
In contrast, the energy contribution from the remaining layers is negligible, characterized by a mix of compute-dominant behavior in GELU and memory-dominant behavior in LN1 and LN2 normalization layers.

\begin{figure*}[t]
    \centering
    \subfloat[\label{fig:resnet50_accuracy}]{%
        \includegraphics[width=0.3\textwidth]{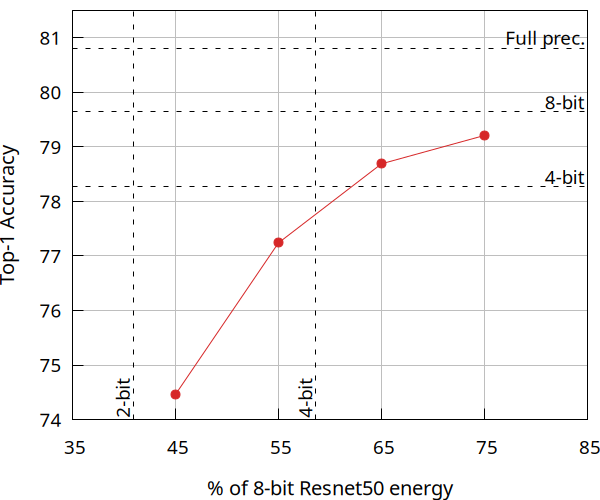}
    }
    \hfill
    \subfloat[\label{fig:resnet50_edp}]{%
        \includegraphics[width=0.3\textwidth]{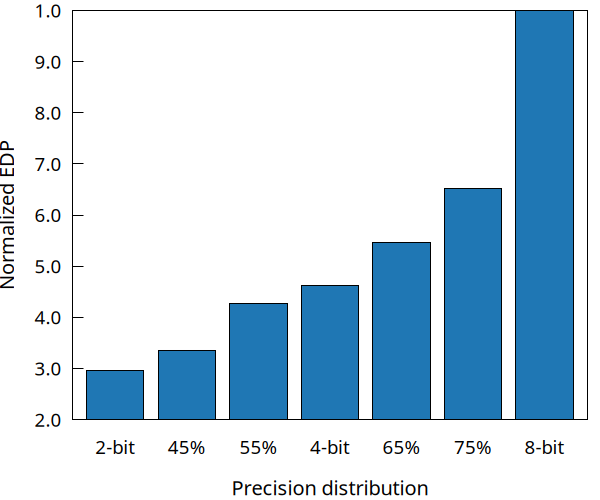}
    }
    \hfill
    \subfloat[\label{fig:resnet50_breakdown_energy_e2e}]{%
        \includegraphics[width=0.3\textwidth]{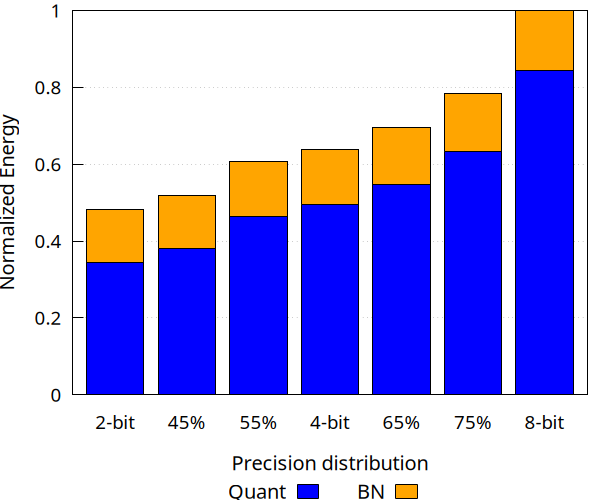}
    }
    \caption{ResNet-50: (a) Achieved fined-tuned accuracy for different energy budgets with respect to an uniform 8-bit implementation of the accelerated kernels; (b) Normalized EDP of comprehensive model for different distributions of layers' precision; (c) Normalized comprehensive energy broken down in terms of the layer type.}
    \label{fig:resnet50}
\end{figure*}

\begin{figure}[t]
 \centering
 \includegraphics[width=0.7\columnwidth]{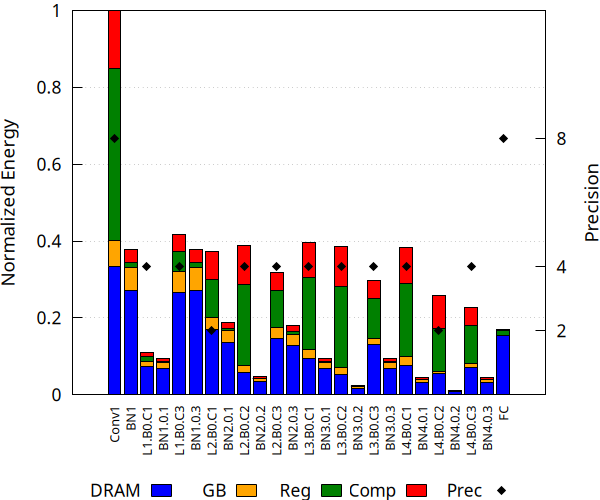}
  \caption{Normalized comprehensive energy broken down for architectural components, for selected ResNet-50 layers.
 Results are normalized to the energy consumption of the initial \textit{Conv1} layer. Precision values are reported exclusively for the accelerated layers of the model.}
  \label{fig:resnet50_breakdown_energy_layer_grouped}
\end{figure}

\subsubsection{ResNet-50}
Figure~\ref{fig:resnet50_accuracy} illustrates the results of our methodology applied to ResNet-50. For energy budgets exceeding 55\%, the accuracy loss remains within 4.5\% of the full-precision baseline.

The comprehensive EDP comparison is presented in Figure~\ref{fig:resnet50_edp}. 
Compared to the 8-bit baseline, EDP reductions range from 35\% for the 75\% mixed-precision solution to 66\% for the 45\% configuration. 
The 55\% mixed-precision solution achieves a 57\% reduction in EDP relative to the 8-bit implementation and a 98\% reduction compared to the full-precision model.

Figure~\ref{fig:resnet50_breakdown_energy_e2e} shows the normalized comprehensive energy for various precision distributions, categorized into accelerated low-precision matrix multiplications (Quant) and BatchNorm (BN) kernels.
Energy savings relative to the 8-bit baseline range from 21.5\% at the 75\% budget to 48\% at the 45\% budget.
For the 55\% mixed-precision network, overall energy is reduced by 39.4\% compared to the 8-bit baseline and by 93\% compared to full precision.

The per-layer energy breakdown for the 55\% mixed-precision ResNet-50, including convolutional (L\textit{x}.B\textit{y}.C\textit{z}, as Layer \textit{x}, Block \textit{y}, Convolution \textit{z}) and subsequent Batch Normalization (BN\textit{x}.\textit{y}.\textit{z}) layers, is detailed in Figure~\ref{fig:resnet50_breakdown_energy_layer_grouped}.
These results reflect the interplay between input feature map size, weight tensor dimensions, the Eyeriss row-stationary dataflow reuse, and the assigned layer precision.
The elevated energy consumption of the initial layer, \textit{Conv1}, stems from a high volume of 8-bit FMA operations and mapping inefficiencies caused by the small input channel dimension. 
Conversely, the energy profile of BatchNorm layers directly correlates with the size of the input hidden vectors.

\begin{table}[ht]
\renewcommand{\arraystretch}{1}
\centering
\caption{Average execution time of SEADA components.}\label{tab:time}
\begin{tabular}{|l|c|c|c|}
\hline
 & BERT-base & ResNet-50 \\
\hline
Generation mapping library & 15 min 40s & 19 min 8s \\
Precision selection & 1 min 42s & 53s \\
Number of epochs for checkpoints & 60 & 40 \\
Number fine-tuning epochs & 60 & 40 \\
\hline
\end{tabular}
\end{table}

\subsubsection{Execution time}
We measured the average execution time of SEADA’s components on an AMD Epyc 7302 @ 3 GHz, utilizing 8-thread parallel execution for the mapping phase. To optimize the creation of the mapping library, we performed the search for optimal mappings only once for all layers sharing identical characteristics.
Table~\ref{tab:time} details these timing measurements, categorized into library generation, the mixed-precision selection process (encompassing empirical entropy computation, Knapsack optimization, and checkpoint generation) and the total number of fine-tuning epochs required.
Due to a greater variety of unique layer configurations, the overhead for generating the ResNet-50 mapping library is generally higher than for BERT.
The cumulative time for library generation and network selection remains negligible compared to the overall training and fine-tuning duration.

\subsubsection{Cost of data conversion}
We compared full-precision layers with and without fused quantization/de-quantization steps. 
Fusing these operations masks their overhead and reduces memory-related energy as full-precision layers write quantized operands back to memory instead of full-precision values.
Estimates show a 2\% energy reduction for mixed-precision BERT-base (45\% solution) versus 40\% for ResNet-50 (55\% solution). 
This disparity occurs because BERT’s energy is dominated by compute-intensive kernels where memory savings are negligible. 
Conversely, ResNet-50 features memory-bound BatchNorm layers with low arithmetic intensity, making it highly sensitive to memory access efficiency.

%% file: sections/dse.tex
\section{Architectural design-space exploration}
\label{sec:dse}

\begin{figure}[t]
 \centering
 \includegraphics[width=0.7\columnwidth]{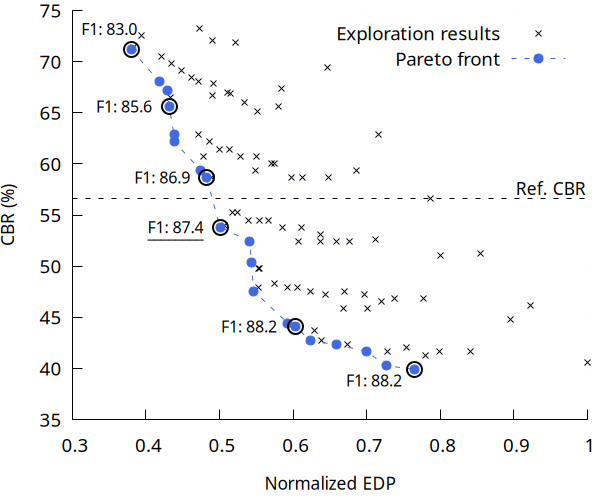}
 \caption{Results of the design space exploration. The Pareto frontier consists of configurations that jointly minimize the EDP and the CBR.}
  \label{fig:dse}
  \vspace{-0.5cm}
\end{figure}

SEADA’s high efficiency enables its application within a more comprehensive architectural design-space exploration.
As a case study, this section investigates the optimal configuration of the baseline accelerator’s memory hierarchy.

In previous experiments, we explored various hardware budgets to derive that 45\% mixed-precision configuration maximized energy savings with minimal accuracy degradation.
As architectural parameters vary, the relationship between mixed-precision distributions and accuracy degradation shifts for each configuration.
Unlike high-level proxy metrics such as bit Multiply-Accumulate (bMAC) counts, the accuracy achieved for a given hardware budget is intrinsically linked to the mapping strategy, the optimization metric, and hardware-specific implementation details.

In our architectural design exploration flow, rather than fine tuning every explored configuration to determine its specific accuracy, we adopted the Computational Bit Reduction (CBR) metric, defined as the percentage reduction in the total number of bMACs between a uniform 8-bit baseline and a given mixed-precision model.
This serves as a hardware-independent proxy for post-fine-tuning accuracy; intuitively, a higher CBR increases the likelihood of significant accuracy degradation.

Assuming an EAGL-like approach can determine the optimal precision distribution under a constrained hardware budget, it is critical to emphasize that the relationship between CBR and accuracy is an intrinsic property of the DNN model, governed by the entropy of its constituent layers.
Once this relationship is characterized for a baseline architecture, the CBR metric serves as a hardware-agnostic proxy for accuracy across all explored architectural configurations.

As the 45\% mixed-precision configuration was identified as the operating point that achieves the highest energy savings while constraining the accuracy degradation to within 1\%,  we adopted the corresponding CBR as upper bound for the selection of candidate solutions during the design space exploration.

In our exploration, we fixed the total hardware area and the number of PEs and we swept the sizes of the global buffer and all register files (input, weight, and output).
The baseline storage area was determined by Equation~\ref{equ:area}.

{\small
\begin{equation}
Area = \#PE \times Area(3 \times 32B\ RFs) + Area (2MB\ GB)
\label{equ:area}
\end{equation}
}

To better leverage the heterogeneous memory hierarchy, we allow the mapping tool to fully explore the mapping space, rather than constraining it to the row-stationary dataflow originally employed in Eyeriss.
We evaluate this approach on the BERT-base workload, using overall EDP minimization as the primary optimization objective.

The register file size of each PE was varied in increments of 32B.
For each configuration, the corresponding global buffer area was derived using Equation~\ref{equ:area}, with a granularity of 8KB.
For every architectural instance, the overall EDP and CBR were evaluated under energy budgets ranging from 35\% to 60\% of the energy of the corresponding 8-bit configuration, in increments of 5\%.

Figure~\ref{fig:dse} reports the results of the design space exploration, highlighting the Pareto-optimal points that jointly minimize the overall EDP and the CBR.
The horizontal line denotes the CBR associated with the 45\% mixed-precision configuration identified in Section~\ref{sec:res-bert}.

To validate our hypothesis regarding accuracy degradation, the configurations highlighted in the figure were fine tuned, and their corresponding F1 scores were measured.
As shown in Figure~\ref{fig:dse}, the accuracy degrades rapidly for CBR values exceeding the reference threshold, thereby supporting the use of CBR as an efficient proxy metric for accuracy during the design space exploration.

The total exploration time required 6 hours and 49 minutes to complete.
For larger design spaces, heuristics and pruning techniques can be easily integrated to significantly reduce exploration time.

The selected Pareto-optimal configuration underlined in Figure~\ref{fig:dse} utilizes a 1776 KB Global Buffer and 192-byte register files.
It achieves a total EDP and energy reduction of 51\% and 20\%, respectively, compared to a uniform 8-bit model. 
When compared to a full-precision baseline, these savings increase to 97.7\% for EDP and 90\% for energy.

%% file: sections/related.tex
\section{Related Work}
\label{sec:related}
The use of mixed-precision in DNNs, in which a different precision is used for different layers, has been recently proposed as a method to reduce computational costs.
Several methodologies have been proposed to optimally select the precision distribution.

Research has evolved toward treating mixed-precision assignment as a constrained optimization problem, typically aimed at balancing hardware efficiency with model sensitivity.

HAWQ-V3\cite{Yao-2021} utilizes Integer Linear Programming (ILP) and a Hessian-based sensitivity metric\cite{Dong-2020} to optimize bit-precisions, notably using dyadic arithmetic to maintain an integer-only pipeline.
Similarly, \cite{Chen-2021} employs a second-order Taylor expansion to quantify sensitivity, reformulating precision selection into a Multiple Choice Knapsack Problem to maximize accuracy within fixed resource budgets.

Alternative strategies, such as ALPS \cite{Bablani-2024}, rank layer importance through empirical accuracy drops, while EAGL \cite{Bablani-2024} introduces a faster analytical approach using weight distribution entropy as a proxy for sensitivity. 
By framing the selection as a 0-1 Knapsack Problem, EAGL efficiently prioritizes higher bit-widths for high-entropy layers, ensuring optimal compression with minimal accuracy loss.

Our work extends the EAGL methodology to support a broader range of precision choices while addressing the limitations of abstract cost models.
Unlike previous approaches that rely on simplified metrics, we incorporate a high-fidelity analytical model that reflects specific architectural details of spatial accelerators.
This model accounts for optimized application mapping, the overhead of quantization and de-quantization, and the energy consumption of non-quantized kernels, providing a more accurate representation of system-level performance.

Research related to our approach can be also found in the literature focusing on the mapping of DNN algorithms onto spatial architectures to minimize specific cost metrics. 
Tools such as MAESTRO \cite{Kwon-2019} and Timeloop \cite{Parashar-2019} utilize the dataflow concept to represent spatial accelerator mappings and assess the performance of CNN algorithms.
These tools employ activity-based analytical models to evaluate the computational and memory costs associated with selected mapping design points.
Additionally, exploration tools like GAMMA \cite{Kao-2020} facilitate the investigation of per-layer dataflows optimized for specific hardware accelerators, aiming to minimize latency, energy consumption, or EDP. 
Other tools, including LOMA \cite{Symons-2021}, CoSA \cite{Huang-2021}, and SALSA \cite{Jung-2023}, focus on the optimal mapping and scheduling of DNNs onto spatial accelerators. 
They achieve this through various methods, including brute-force exploration, mixed-integer linear programming formulations, and simulated annealing to optimize loop structures and orders.
FactorFlow \cite{Ronzani-2025.1} and QuickFlow \cite{Ronzani-2025.2} further refine the mapping design space for GEMMs and CNNs, respectively, by leveraging adaptive programming and greedy optimization techniques, achieving local optimal solutions with exploration times that are an order of magnitude faster than those of earlier tools.

The above mentioned frameworks support the exploration of the DNN mapping space but assume a single, generic, and unspecified precision for computations.
How framework leverages the exploration capabilities of QuickFlow incorporating an activity-based cost model for multi-precision architectures, allowing expanding the exploration of the mapping of mixed-precision DNNs on spatial architectures.

The QUIDAM framework \cite{Inci-2023} enables a unified design-space exploration by simultaneously optimizing DNN models and hardware accelerators across various bit-precisions.
A key finding of the study is that the interplay between bit-precision and PE creates an expansive design space, where performance-per-area and energy efficiency can vary by orders of magnitude. 
By leveraging low-precision lightweight PEs, the researchers achieved significantly higher hardware efficiency than traditional INT16-based implementations without compromising model accuracy.

Although QUIDAM models low-precision pipelines, its optimization focus is limited to single-precision processing, offering no support for mixed-precision implementation exploration.

In \cite{Klhufek-2024}, a joint optimization framework is introduced that couples mixed-precision quantization with hardware mapping.
By enhancing Timeloop to handle variable bit-widths, the authors demonstrate how reduced precision improves data tiling and bit-packing.
The search space is navigated using the NSGA-II algorithm, balancing accuracy against energy and memory constraints.
A notable constraint of this work is that while it optimizes data movement and storage through mixed precision, the underlying MAC units remain at a fixed precision and are not part of the optimization process.

The work in ~\cite{Fiorin-2025} pioneered the integration of mixed-precision into dataflow exploration for spatial architectures.
In SEADA, we introduce a redesigned framework built upon Quickflow, significantly expanding the analytical reach of~\cite{Fiorin-2025}.
Key enhancements include the integration of dedicated cost models for quantization/de-quantization and un-quantized kernels, alongside a refined methodology for per-layer precision selection and the rapid evaluation of inference accuracy.
By co-optimizing these hardware and software variables, SEADA provides a more holistic view of the trade-offs in mixed-precision DNN deployment on spatial hardware.

%% file: sections/conclusions.tex
\section{Conclusions}
\label{sec:conclusions}
This work presents SEADA, a fast methodology for the optimization of mixed-precision DNNs on multi-precision architectures. 
SEADA comprises:
(i)  a configurable system-level analytical cost model of a multi-precision spatial accelerator architecture;
(ii) a fast mapping tool that identifies near-optimal mappings of DNN workloads onto the target integer accelerator;
(iii) analytical models for floating-point layers to estimate the overall benefits of mixed-precision execution;
and (iv) a per-layer precision selection methodology based on bit-level entropy, enabling efficient assignment across multiple numerical precisions.

SEADA’s high efficiency enables its application within a more comprehensive architectural design-space exploration.
We investigated the optimal configuration of a multi-precision accelerator’s memory hierarchy, demonstrating how a novel metric, Computational Bit Reduction, serves as a reliable proxy for model accuracy. 
This approach allows for exploration without the need to fine tune the DNN model for every architectural configuration.

Upon acceptance of this article, the complete source code will be released as open-source to support reproducibility and further research.